# Sieving and clogging in PEG-PEGDA hydrogel membranes

Malak Alaa Eddine[1,2], Alain Carvalho[3], Marc Schmutz[3], Thomas Salez[4], Sixtine de Chateauneuf-Randon[1], Bruno Bresson[1], Sabrina Belbekhouche*[2], Cécile Monteux*[1]

1- Laboratoire Sciences et Ingénierie de la Matière Molle, ESPCI Paris, 10 rue Vauquelin, Cedex 05 75231 Paris, France.
2- Université Paris Est Creteil, CNRS, Institut Chimie et Matériaux Paris Est, UMR 7182, 2 Rue Henri Dunant, 94320 Thiais, France.
3- Université de Strasbourg, CNRS, Institut Charles Sadron, 23 rue du Loess, 67034 Strasbourg Cedex 02, France.
4- Univ. Bordeaux, CNRS, LOMA, UMR 5798, F-33400 Talence, France.

* Authors for correspondence:
E-mail addresses: cecile.monteux@espci.fr (C. Monteux).

belbekhouche@cnrs.fr (S. Belbekhouche).

**Abstract**

Hydrogels are promising systems for separation applications due to their structural characteristics (i.e. hydrophilicity and porosity). In our study, we investigate the permeation of suspensions of rigid latex particles of different sizes through free-standing hydrogel membranes prepared by photopolymerization of a mixture of poly (ethylene glycol) diacrylate (PEGDA) and large poly (ethylene glycol) (PEG) chains of 300 000 g.mol$^{-1}$ in the presence of a photoinitiator. Atomic force microscopy (AFM) and cryoscanning electron microscopy (cryoSEM) were employed to characterize the structure of the hydrogel membranes. We find that the 20 nm particle permeation depends on both the PEGDA/PEG composition and the pressure applied during filtration. In contrast, we do not measure a significant permeation of the 100 nm and 1 µm particles, despite the presence of large cavities of 1 µm evidenced by cryoSEM images. We suggest that the PEG chains induce local nanoscale defects in the cross-linking of PEGDA-rich walls separating the micron size cavities, that control the permeation of particles and water. Moreover, we discuss the decline of the permeation flux observed in the presence of latex particles, compared to that of pure water. We suggest that a thin layer of particles forms on the surface of the hydrogels.

**Keywords:** Hydrogel, poly (ethylene glycol) diacrylate, poly (ethylene glycol), latex particles, AFM, cryoSEM, clogging, filtration, porous flows.

**Introduction**

Hydrogels are networks of polymer chains swollen in water in which molecular species and nanoparticles can diffuse under an imposed concentration gradient or can be transported through a hydrodynamic flux obtained by applying a pressure gradient[1]. Inversely, the molecular network structure of hydrogels provides them with promising filtration properties [2]. The diffusive transport of species in hydrogels has been the object of numerous experimental and theoretical studies in the past, often motivated by drug release applications [3-4]. Several theories have been developed such as the obstruction, hydrodynamic or free volume theories with various assumptions. While the obstruction theory based on the Ogston model assumes that the hydrogel is a rigid network with a fixed mesh size [5], other theories take into account hydrodynamic interactions between the network and the solvent or the thermal fluctuations of the network [6-7]. Several experimental studies provided measurements of diffusion coefficients of molecules in hydrogels and compared them with the available theories [8]. Overall, the size of the solute, its charge and its interactions with the components of the gel, control its transport through a hydrogel membrane [2, 9-11]. While various studies exist on the particle transport through porous filtration membrane systems under the effect of applied pressure gradients [12-15], the transport of particles through hydrogel membranes is relatively poorly studied [16].

The use of hydrogels as coating layers on classical hydrophobic filtration membranes has attracted attention for wastewater treatment [17-19]. They have proven their effectiveness as ideal materials for modifying filtration membranes, to achieve robust anti-fouling and long stability [19-21]. However, coating or grafting hydrogels on membranes still presents some issues, mainly by the partial adsorption of the pollutants that pass through the hydrogel layer on the hydrophobic substrate during filtration [22-23]. A great interest has been paid in recent years to self-supporting hydrogel films with a thickness of several hundreds of microns, in order to avoid surface-coated membrane problems [24-25]. Free-standing hydrogels have been used for water permeability studies and the rejection of dyes, proteins and salts [26-27]. Since common hydrogels have poor mechanical properties [28], it is important to develop new hydrogel systems that could be used as free-standing membranes without support .

Poly (ethylene glycol) diacrylate (PEGDA)-based hydrogels are resistant hydrogels which have been widely used as scaffolds for tissue engineering applications [29-30], drug delivery [31] and microfluidic devices [32-33]. In the context of filtration, PEGDA hydrogels have proven their effectiveness as protein-fouling resistant films [31, 34]. In addition, they have shown a well-controlled water permeability through the porous PEGDA films complemented with porogens (e.g. poly (ethylene glycol) PEG chains with low molar mass) [35-36]. PEG-based hydrogel membranes are known for their size-dependent particle permeabilities, and used as simple models for mucus [1, 37-38]. Recently, we have developed new PEGDA/PEG composite hydrogel membranes by introducing large free PEG-300 000 g.mol$^{-1}$ chains to a PEGDA matrix [39]. We have shown that these large PEG chains remain trapped in the hydrogel matrix during multiple filtration cycles. Using a frontal filtration cell, we were able to show that the water permeability, $K$, varies over two orders of magnitude depending on the PEG concentration (from $1 \times 10^{-18}$ to $5.5 \times 10^{-16}$ m$^2$) and presents a maximum with the PEG concentration corresponding to $C^*$, the overlap concentration.

In the present study, we investigate the filtration of rigid latex particles with different sizes (20 nm, 100 nm and 1μm) through these PEGDA/PEG hydrogel membranes. We focus on how both water and the particles permeate through the hydrogels depending on pressure and hydrogel composition. For the suspensions, the water flow rate is reduced possibly because of particles clogging the surface of the hydrogel. Moreover, the permeation of the particles depends on the applied pressure in a non-trivial manner possibly because of the deformability of the hydrogel. Finally, we combine these particle permeation results with AFM and cryoSEM results to obtain more insight into the hydrogel structure. Despite the fact that all PEG containing samples present micron sized cavities, particles of 100 nm and 1 μm do not permeate through the membranes. We suggest therefore that these micron size cavities do not form a percolating network and that the permeation of water and nanoparticles is rather controlled by the nanostructure present in the PEGDA rich walls between these micron size cavities. This suggestion is consistent with a previous study on PEGDA/PEG hydrogels [39].

**Experimental section**

**Materials**

We used poly (ethylene glycol) diacrylate PEGDA ($\overline{Mw} = 700$ g.mol$^{-1}$) with 13 ethylene oxide units and 4-(2-hydroxyethoxy) phenyl 2-hydroxy-2-propyl ketone (Irgacure 2959) which were purchased from Sigma–Aldrich. Linear poly (ethylene glycol) (PEG) ($\overline{Mw}$ =300 000 g.mol$^{-1}$, polydispersity index Đ = $\overline{Mw}/\overline{Mn}$=2.1) were purchased from Serva. FluoSpheres™ carboxylate-modified polystyrene particles ((20 nm, 100 nm and 1 µm of diameter), red fluorescent (580/605), 2% solids corresponding to a concentration of 2x10$^{-2}$ g.mL$^{-1}$) were purchased from thermoFisher scientific. In Supporting information S1 is provided the DLS measurements of the particles diameter as well as their zeta potential. Water was purified with a Milli-Q reagent system (Millipore).

**PEGDA hydrogels preparation**

The PEGDA and PEG/PEGDA membranes were synthesized via UV-initiated free-radical photopolymerization using Irgacure 2959 as the photoinitiator. The exact compositions are given in the Supporting Information Table S2. Briefly, the prepolymerization solution were composed of a fixed mass of PEGDA (2 g), Irgacure (2 mg) and water (20 g) and increasing amounts of PEG 300 000 g.mol$^{-1}$ (between 0.05 g and 0.5 g). We choose this recipe so as to keep the PEGDA/water ratio constant and equal to 0.2 as this parameter was shown to play an important role on PEGDA hydrogels structure [40]. Moreover, the Irgacure/PEGDA ratio was fixed to 0.1 wt% similarly to several values of the literature [41]. The wt% of PEG varies between 0.4 and 4 wt%. In Supporting Information S2 is provided a Table with all the compositions used.

To prepare the hydrogel membranes, the prepolymerization solutions were sandwiched between two glass plates (120 mm x 80 mm) which were separated by 1 mm thick spacers to obtain membranes of thickness of 1 mm. Then the solution were polymerized under irradiation of UV light (intensity =1800 µw/cm$^2$) with a wavelength of 365 nm for 10 min similarly to other studies [41]. After polymerization, the obtained hydrogel sheets were placed in a petri dish with pure water for at least 24 hours prior filtration to eliminate any unreacted PEGDA monomers or small free PEG chains. As explained in our previous article we found using Total Organic Content measurements, that less than 3% of the PEG chains get removed in the supernatant or during filtration. To prepare the hydrogel membranes, these large hydrogels sheets were then cut with a round die puncher of 45 mm in diameter to obtain 1 mm thick disks of 45 mm diameter.

To follow the polymerization reaction of PEGDA, we have evidenced in our previous work [39] using IR spectroscopy that the characteristic peak of the C=C bonds at 1633 cm$^{-1}$ from PEGDA oligomer has disappeared for the cross-linked PEGDA or mixture of PEGDA/PEG samples. This suggests that the hydrogel films were successfully synthesized after UV cross-linking.

**Particles characterization**

Zeta potential and particles size measurements of carboxylate-modified polystyrene particles (Fluospheres) were recorded by Zetasizer Nano-ZS90 from Malvern. The samples used for these measurements were 100-fold diluted in Milli-Q water to reach a particle concentration of $C_0=2\text{x}10^{-4}$ g.mL$^{-1}$ and to avoid multiple scattering of light. The results are presented in Supporting Information (Table S1 and Figure S1). The optical absorbance measurements of the modified polystyrene particles were carried out with a UV–vis Hewlett-Packard 8453 spectrophotometer using a 1 cm path length quartz cell, in a wavelength range from 190 to 1100 nm.

**AFM characterization**

AFM images were obtained with a Bruker Icon microscope driven by a Nanoscope V controller. The surface of the hydrogel membrane immersed in water before filtration was observed in Peak Force mode. The height images were acquired with a cantilever of spring constant 0.7 N.m$^{-1}$ specially designed for this application. In this mode, similar to a rapid approach-retract experiment, the cantilever oscillates at a frequency of 1 kHz. The scanning frequency was 0.7 Hz and the maximum force was set to 500 pN. We chose to use AFM to image the membrane surface in situ in water.

**CryoSEM characterization**

PEGDA and PEGDA/PEG hydrogel membranes with a thickness of 1 mm were placed on a home-made cryo-holder to be quickly plunged into an ethane slush. As the sample is free-standing over the holder, the sample is rapidly frozen during the plunging by direct contact with the liquid ethane, in order to form amorphous glace. Subsequently, the sample is transferred into the Quorum PT 3010 chamber attached to the microscope. There, the frozen sample is fractured with a razor blade. A slight etching at −90°C may be performed to render the sample more visible. The sample is eventually transferred in the FEG-cryoSEM (Hitachi SU8010) and observed at 1 kV at −150°C. No further metallization step is required before transferring the sample to the SEM chamber. Several sublimation cycles were performed on each sample to ensure the removal of the glace from the hydrogel.

**Particles filtration experiments**

Particle filtration through PEGDA hydrogels and PEG-modified PEGDA hydrogels was measured using a dead-end ultrafiltration (UF) cell obtained from Fisher Scientific S.A.S. (Model 8050, 50 mL for 45 mm Filters) as represented in Figure 1. The filtrations are performed at ambient temperature, with $V_0$=60 mL of carboxylate-modified polystyrene particles (Fluospheres) solution as the feed solution. The solution used for this experiment was 100-fold diluted in Milli-Q water so that the concentration of the feed solution was fixed constant and equal to $C_0$=2x10$^{-4}$ g.mL$^{-1}$. The membrane area was $S_m$=15.90 cm$^2$ fixed in the membrane holder of the cell. Standard filtration experiments were performed at a pressure ranging from 2 KPa to 80 KPa. To obtain in a the time evolution of the flow rate $Q$ in m$^3$.s$^{-1}$ as well as $C_p$, the particle concentration in the permeate, aliquots of permeate, of volumes ranging between 1 and 10 mL, were collected in a continuous manner over time and weighted to obtain $Q$. The latex concentration in these aliquots was determined using a UV–vis spectrophotometer to obtain the instantaneous concentration, from which the total permeate concentration, $C_p$, was calculated by taking into account all the collected volumes and concentrations. The calibration curves for UV-visible measurements are provided in Supporting Information Figure S3-A.

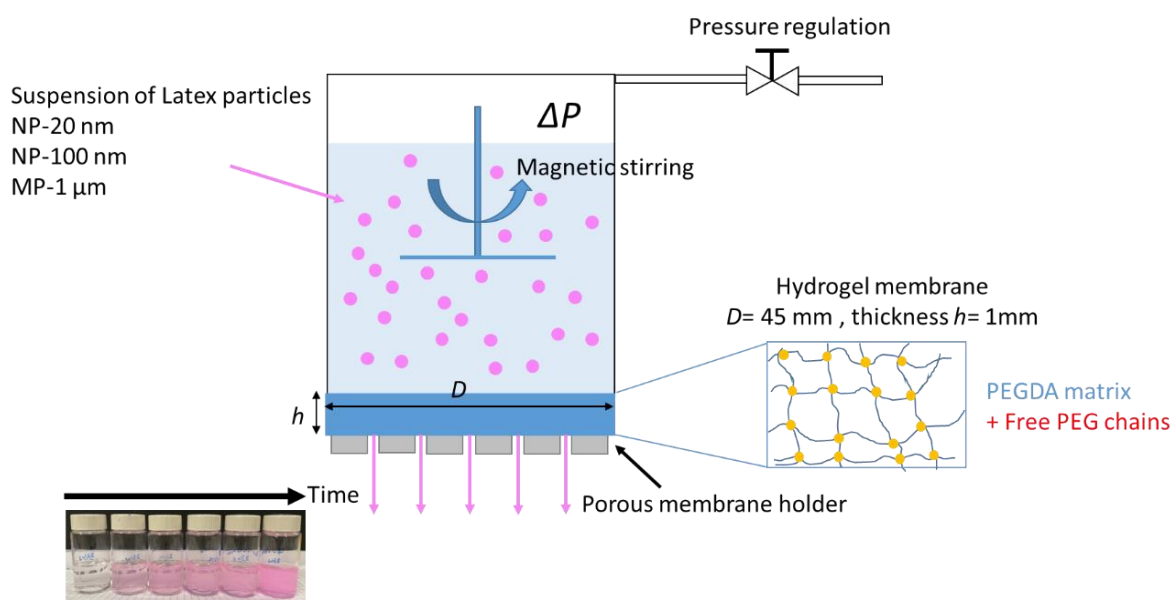

*Figure 1. Representative schematic of the filtration experiments of latex particle suspension using an ultrafiltration stirred cell.*

## RESULTS AND DISCUSSION

**Permeation of suspensions of nano and micrometric particles by PEG/PEGDA hydrogels**

*Water and suspension fluxes through the hydrogels*

We first compare the water permeation flux through the PEG/PEGDA hydrogels when the feed solution is either pure water or suspensions of nano and microparticles. The choice of the hydrogels compositions studied is based on our previous work [39], in which we found that the water permeability, noted $K$, of PEGDA/PEG hydrogels presents an optimum at a weight fraction of PEG of $C^* = 1.6$ wt% corresponding to the overlap concentration of the PEG chains. We therefore choose PEGDA/PEG compositions that range below and above this maximum permeability (i.e. $K = 5.5 \times 10^{-16}$ m$^2$). We compare the case of pure PEGDA hydrogels and PEGDA/PEG hydrogels containing 0.4 wt%, 1.6 wt% and 4 wt% of PEG (Figure 2). In the case of the pure PEGDA samples, the flow rate measured with pure water varies linearly with the pressure as already reported in our previous study [39]. For the PEG containing samples, the permeation flux measured for pure water varies non linearly with the applied pressure and presents a plateau at the highest pressures above 50 KPa. This is consistent with our previous work in which we showed that this phenomenon is due to the high compressibility of the PEGDA/PEG samples inducing a reduction of the porosity at high pressure.

In the case of the suspensions containing the 20 nm particles, the measured fluxes are identical to those obtained for pure water in the case of the pure PEGDA hydrogels and the 4 wt% PEG hydrogels (Figure 2 a and d). However, for the 0.4 wt% (Figure 2 b) and 1.6 wt% PEG (Figure 2 c) hydrogels, the permeation flux of the 20 nm suspension is lower than the one measured for pure water. In Figure 3, we compare the influence of particle size on permeation fluxes of the suspensions for the 1.6 wt% PEG/PEGDA hydrogels. The presence of latex particles decreases the permeate flux for the three types of particles investigated, the largest effect being obtained for the large particles of 100 nm and 1 µm.

Such a reduction of the flux can either be due to a clogging of the hydrogel membrane by the particles or to the formation of a cake of particles on the surface of the hydrogel. We note that the total volume of 20 nm particles in the permeate is $V_{tot}^{20\,nm} = C_0 . V_0 = 0.012$ cm$^3$ which is more than a hundred times lower than the volume of the membrane, $V_m = 1.5$ cm$^3$. For the 100 nm and 1µm particles the number of particles in the feed solution is even lower for the same weight percentage as the particles have a larger volume. Hence we suggest that the reduction of the permeation flux is probably not caused by the clogging of the particles inside the hydrogel. In the case of the gradual formation of a cake, one expects that the permeability of the cake will decrease over time as the cake thickness will increase, and hence we should observe a temporal decay of the permeation flux. However, in Supporting information S4 we show that the permeate volume varies linearly with the time, for all the applied pressures, showing that the permeation flux is constant over time. In fact, the solution is stirred magnetically during the filtration and probably prevents the growth of such a cake. We suggest that the formation of a thin layer of particles blocking some of the pores at the surface of the membranes may account for the flux reduction. Indeed, we can evaluate that the total surface of the particles, $S_{tot}^{20\,nm}$, is much higher than the surface of the membrane, $S_m$, where $S_{tot}^{20\,nm} = \frac{C_0 \cdot V_0 \cdot S_{part}^{20\,nm}}{V_{part}^{20\,nm}} \gg S_m$.

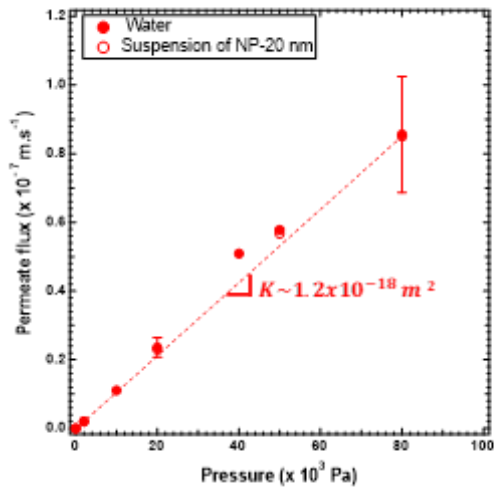

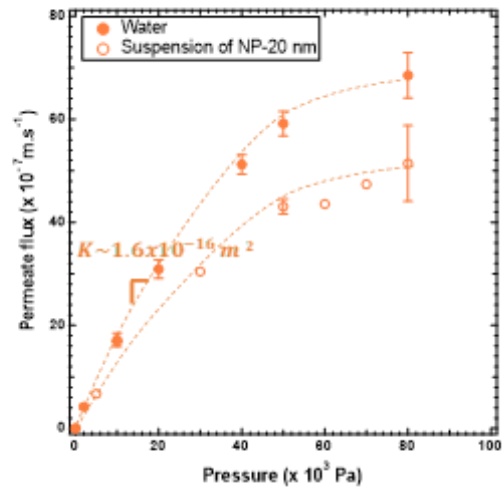

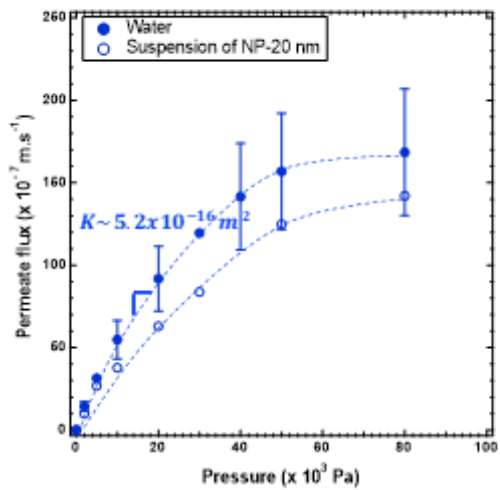

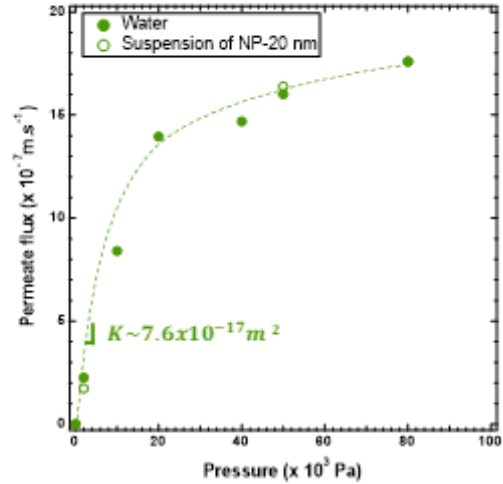
8

*Figure 2. Variation of the water and latex nanoparticles of 20 nm (NP-20 nm) permeate fluxes as a function of applied pressure for hydrogel membranes prepared with PEGDA and a) 0 wt%, b) 0.4 wt%, c) 1.6 wt% and d) 4 wt% of PEG-300 000 g.mol $^{-1}$. The dashed lines are guides for the eye.*

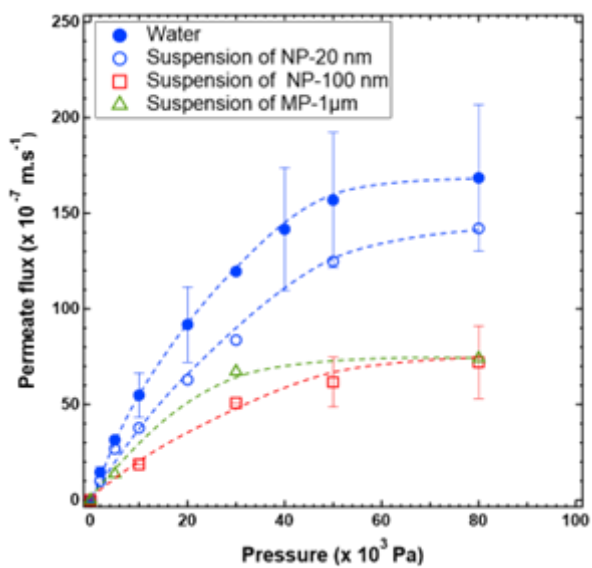

*Figure 3. Variation of the water, latex nanoparticles of 20 nm (NP-20 nm), nanoparticles of 100 nm (NP-100 nm) and microparticles of 1 μm (MP-1μm) permeate flux as a function of applied pressure for hydrogel membranes prepared with PEGDA and 1.6 wt% of PEG-300 000 g.mol$^{-1}$. The dashed lines are guides for the eye.*

We estimate the reversibility of such surface clogging process of the hydrogels by the particles by performing filtration cycles, increasing and then reducing the pressure and measuring the permeation flux (Figure S5). In the case of the 100 nm particles, after a first increase of the pressure to 80 KPa, decreasing the pressure back to lower values does not enable it to recover the same value of the permeation flux as the one measured as the pressure is increased. This means that the clogging of the cavities by the 100 nm is irreversible. Indeed, we obtained cryoSEM images of the hydrogels after the filtration of the 100 nm (Figure S6a) and find that the 100 nm particles are indeed trapped in the top part of the hydrogels. At opposite, the cyclic filtration of MP-1μm through the hydrogel does not significantly affect the permeate flux and the particles were not observed inside the hydrogels after filtration (Figure S6b). This result confirms that these micron sized particles do not enter into the hydrogel and that the flux reduction is due to the reversible deposition of a thin layer of particles of constant thickness on the hydrogel membrane blocking some of the surface pores.

*Time evolution of the particles permeation through the hydrogels*

In order to determine how the particles permeate through the hydrogels, we measure the time evolution of the concentration of 20 nm particles in the permeate for different applied pressures and hydrogel compositions. For pure PEGDA hydrogel (Figure 4 a), we find that the concentration of nanoparticles of 20 nm (NP-20 nm) does not increase significantly over time hence the particles do not permeate through the hydrogels. For the 0.4 wt% PEG samples (Figure 4 b) the concentration in the permeate, $C_p$, tends to increase almost linearly with time, with a higher slope as the pressure is increased. For 80 KPa, the concentration tends to saturate to the initial concentration of the feed solution, $C_0$, after approximately 80 minutes. For a pressure of 5 KPa, the concentration of nanoparticles in the permeate does not increase significantly over time meaning that the particles do not permeate across the hydrogels. We will discuss the influence of pressure in the permeation of the particles below. Similarly, for the most permeable PEGDA hydrogel containing 1.6 wt% of PEG (Figure 4 c), the concentration of NP in the permeate increases more rapidly as the pressure is increased. For 2 KPa, permeate concentration does not increase significantly.

In the case of 4 wt% of PEG content in the hydrogel (Figure 4 d), we find that the concentrations of NP-20 nm in the permeates are an order of magnitude below the values observed for the 0.4 and 1.6 wt% of PEG/PEGDA membranes and are independent of the applied pressure and the permeate volume (even for the largest applied pressure of 80 KPa).

To better understand the influence of the applied pressure on the time evolution of the particle permeation, we must take into account the fact that the pressure influences the flow rate of the suspension through the hydrogels, hence the velocity of both water molecules and particles in the hydrogels. To take this effect into account, we plot the curves of Figure 4b and 4c as a function of $= \frac{t.Q}{V_m}$, where $V_m = 1.5 \times 10^{-6}$ m$^3$ is the volume of the hydrogel membrane, and $Q$ (in m$^3$.s$^{-1}$) is the suspension flow rate corresponding to $J.S_m$, where $J$ is the permeation flux (in m.s$^{-1}$) whose values are presented in Figures 2 b and 2 c. and $S_m$ is the surface of the membrane. We note that $\lambda$ can also be written as $= \frac{t}{\tau} = \frac{V_p}{V_m}$, where $\tau$ represents the time required for a water molecule to flow through the thickness ($e$=1mm) of the hydrogel and $V_p$ is the eluted volume. Using such a rescaling enables us to take into account the variation of permeation rate with the pressure and one expects that the curves obtained at various pressures should collapse on a single master curve if the influence of pressure was only to control the permeation velocity. However, Figures 5a and 5b do not rescale on the master curve. For the lowest pressures, 5 KPa and 2 KPa for the 0.4 wt% and 1.6 wt% samples respectively the permeation of the 20 nm particles is almost no detectable even for values of $\lambda$ up to 10 or 20. For intermediate pressures of 10; 50 and 70 KPa, the permeation is significantly higher but does not seem to depend significantly on the pressure. For the largest pressure of 80 KPa the permeation of the particles seems to increase significantly. We suggest that these non-trivial influence of the pressure on the permeation of the particles is due to the softness of the PEGDA/PEG hydrogels and possible deformation of the pores as the particles are forced through the pores upon increasing pressures. This point will be discussed in the next Section.

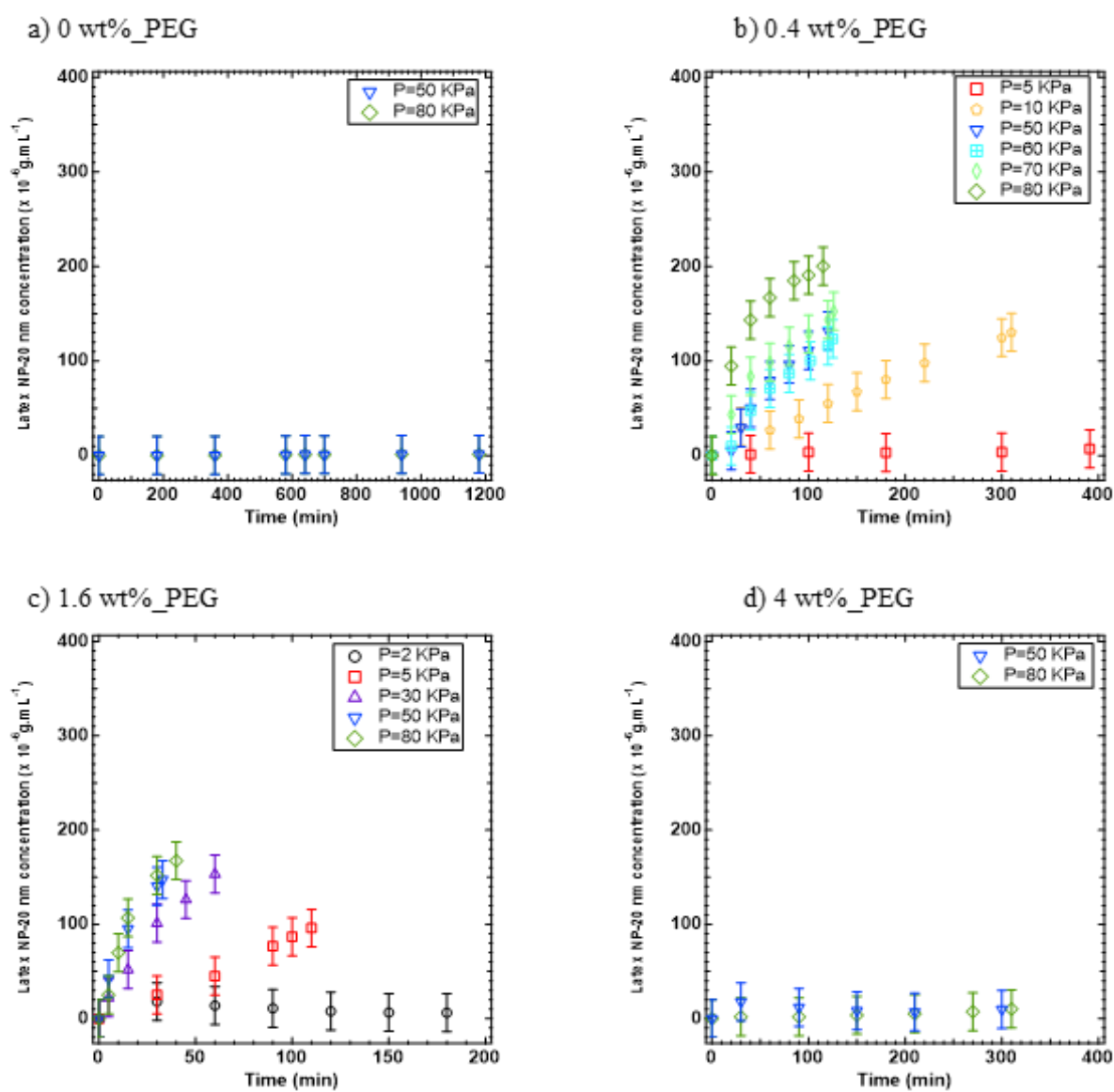

*Figure 4. Variation of latex nanoparticle of 20 nm concentration in the permeate as a function of the time under different applied pressures for hydrogel membrane prepared with PEGDA and a) 0 wt%, b) 0.4 wt%, c) 1.6 wt% and d) 4 wt% of PEG-300 000 g.mol$^{-1}$.*

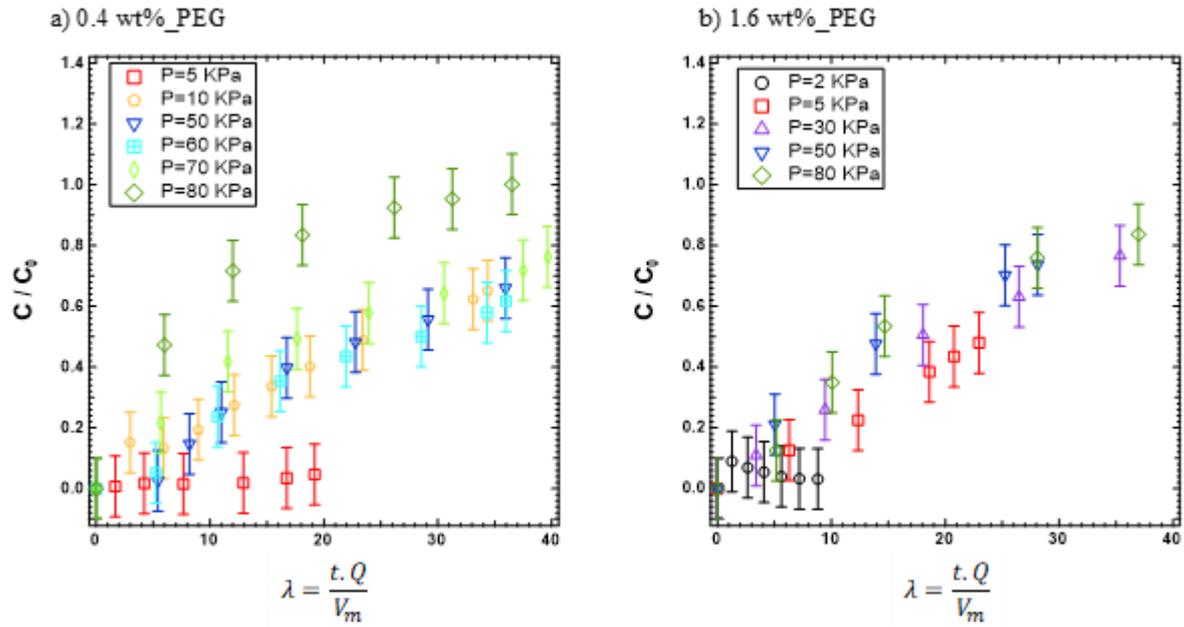

*Figure 5. Rescaling the curves of Figure 4 b and 4c to take into account for the variations of flow rate induced by the pressure variations*

We investigate the influence of particle size and found that for the 1.6 wt% PEG hydrogel (Figure 6), the concentration of 100 nm and 1 µm particles in the permeate remains below the detection threshold confirming that these particles do not permeate through the entire thickness of the hydrogel as already concluded earlier when discussing the values of the permeation fluxes.

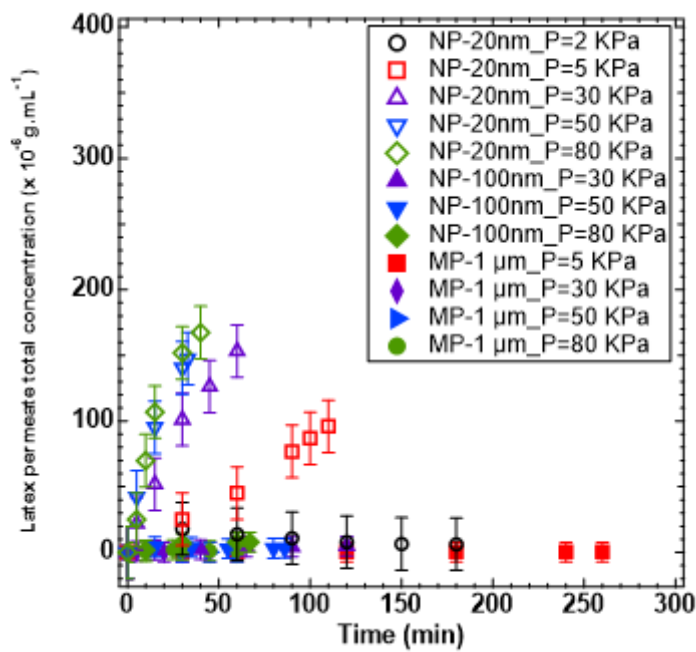

*Figure 6. Variation of latex nanoparticles of 100 nm (NP-100 nm) and microparticles of 1 µm (MP-1µm) concentration in the permeate as a function of the time under different applied pressures for hydrogel membranes prepared with PEGDA and 1.6 wt% of PEG-300 000 g.mol$^{-1}$. The empty symbols correspond to the concentration of latex nanoparticles of 20 nm (NP-20 nm).*

**Link between permeation and structure of the PEG/PEGDA hydrogels**

In this section, we present AFM and CryoSEM results to obtain more insight into the structure of the hydrogels. As can be seen in Figure 7, the cryoSEM images of the PEGDA hydrogels with no added PEG, evidence the presence of cavities of diameter ~ 200 nm (indicated by white arrows). Molina *et al.*[40] used SANS and showed that these large cavities are filled with water. Furthermore, they showed that the mesh size of the PEGDA 700 g.mol$^{-1}$/water matrix is of the order of 1 nm and its equilibrium water content of PEGDA hydrogel is equal to 50 wt% and controlled as in usual hydrogels by a competition between polymer water affinity and loss of entropy of the polymer chains as they stretch because of water swelling. Hence for prepolymerization solutions containing more than 50 wt% of water, there is an excess of water that cannot be incorporated in the polymerizing network and forms 100-200 nm water cavities trapped in the PEGDA matrix. In order to study with more accuracy the structure of the hydrogels, AFM images of the membrane surface were acquired in water with a field of view of 5 µm, allowing us to focus on the nanometric structures. For reasons of image clarity, the Z-scales are not identical for all images and we have chosen to adapt the Z scale at the best for each image. The AFM images confirm the presence of cavities of diameter ~ 200 nm for the pure PEGDA hydrogels.

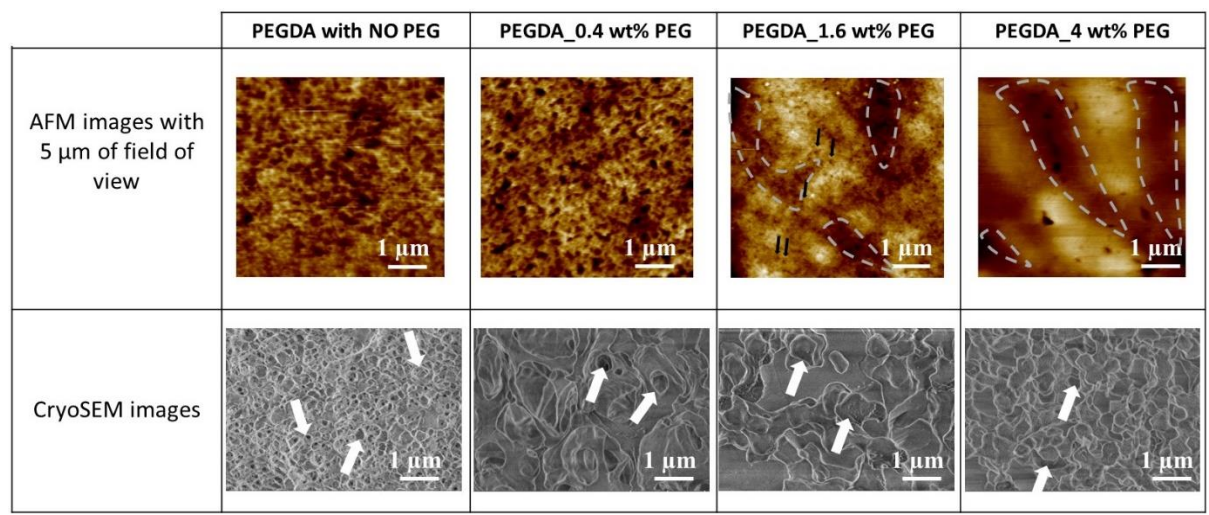

*Figure 7. AFM images and cryoSEM images for hydrogel membranes prepared with PEGDA and various contents of PEG-300 000 g.mol$^{-1}$. The typical Z-scale in AFM images is ± 50 nm – white arrows represent large cavities observed with cryoSEM (200 nm and above); The dotted lines represent micro size heterogeneities observed with AGM and black arrows represente 20 nm heterogeneities observed by AFM*

For the three PEG containing samples investigated, the cryoSEM images show larger heterogeneities of size ~ µm (indicated by white arrows). For the 1.6 wt% sample, AFM confirms the presence of micron scale heterogeneities (dotted lines) as well as smaller cavities of a few tens of nanometers (indicated by black arrows). These 20 nm cavities are not visible for the 0 wt% and 0.4 wt% samples. In fact, for the 4 wt% PEG sample, the walls of the micron size cavities seem to be much more homogenous than for the other compositions.

Let us now discuss a possible structure for the hydrogels that may account for all our results. For the pure PEGDA samples, the rejection of the 20 nm particles during filtration suggests that the 200 nm water cavities observed with cryoSEM do not percolate through the hydrogels. Moreover, we suggest that the permeability of the PEGDA sample is controlled by the 1 nm mesh size of the PEGDA/water with a network of 50wt%/50wt%. From reference [40], we know that the volume fraction of these cavities is of the order of 60 wt%, hence below the random close packing conformation and it is consistent with the unconnected porosity scenario that we propose.

For the PEGDA/PEG samples, let us first discuss the case of the most permeable sample containing the 1.6 wt% PEG. The 20 nm particles permeate through the hydrogels for pressures above 5 KPa while the 100 nm particles do not permeate significantly through the hydrogels for any pressure applied. We suggest therefore that the pore size probably ranges between 20 nm and 100 nm. Moreover, the permeation results showed that for P=2 KPa, the permeation of the 20 nm is not significant. As the hydrogels are soft deformable materials, we suggest that a threshold pressure of 5 KPa is required to cause the deformation of the pores that allows the permeation of the 20 nm particles.

The pressure ($P$) needed to force a particle of radius $R_{part}$ to go through a cavity of radius $R_{cavity}$ in a material of elastic modulus ($E$) can be roughly estimated as (Equation 1):

$$\frac{P}{E} \approx \frac{R_{part}}{R_{cavity}} - 1 \qquad (1)$$

This equation can be rewritten as:

$$\frac{R_{part}}{R_{cavity}} \approx \frac{P}{E} + 1 \qquad (2)$$

The Schematic representation of the model is given in Figure 8.

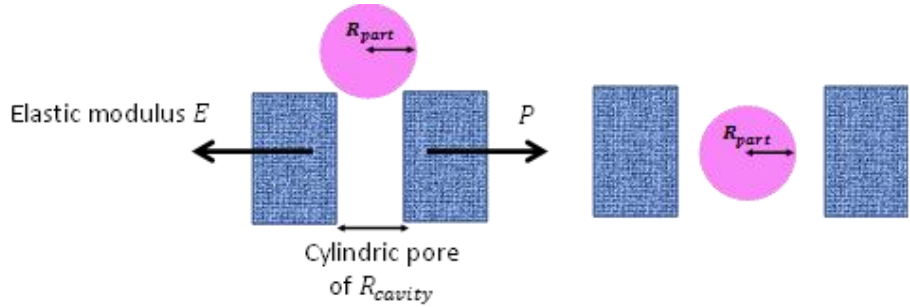

*Figure 8. Schematic representation of the simplified model based on the assumption that the particle is spherical, the pores are cylindrical and it assumes that the pores are uniformly distributed with a constant radius within a porous medium presenting a constant elastic modulus.*

In the case of the 1.6 wt% PEG sample, a low pressure of 5 KPa is required to force the particles to permeate through the hydrogel. Interestingly this low pressure is much lower than the elastic modulus of the PEGDA/PEG hydrogels, of the order of 0.2 MPa [39]. Consequently, one has $\frac{P}{E} \ll 1$ meaning that $R_{cavity} \approx R_{part} \approx 20$ nm. This estimated value $R_{cavity}$ is orders of magnitude lower than $d_{cavity} \sim$ µm observed in the cryoSEM images, It is, however, the same order of magnitude as the heterogeneities observed by AFM in the walls of the micrometric cavities for the 1.6 wt% PEG samples. Therefore, this result suggests that the permeation of the 20 nm nanoparticles through the hydrogel membranes is probably controlled by heterogeneities of the order of 20 nm in the walls connecting the micron sized cavities.

These nanometric heterogeneities in the walls of the micron sized cavities probably also control the water permeation in the hydrogels. We know from a previous study that the PEG chains remain trapped within the hydrogels. Hence, as shown in Figure 9, we suggest that the PEG chains get trapped inside the PEGDA rich walls separating the micron size cavities during the polymerization process. The PEG chains would then induce local defects in the cross-linking density allowing the permeation of the 20 nm nanoparticles but not the 100 nm ones. In such a scenario, the PEG-containing walls would therefore provide some level of connectivity between the large micron cavities and control the permeation of both the particles and water.

Above $C^*$, the permeation of both water and the 20 nm decreases even though the size of the large cavities remain unchanged. We hypothesize that the free PEG and PEGDA form an interpenetrated network with a lower mesh size.

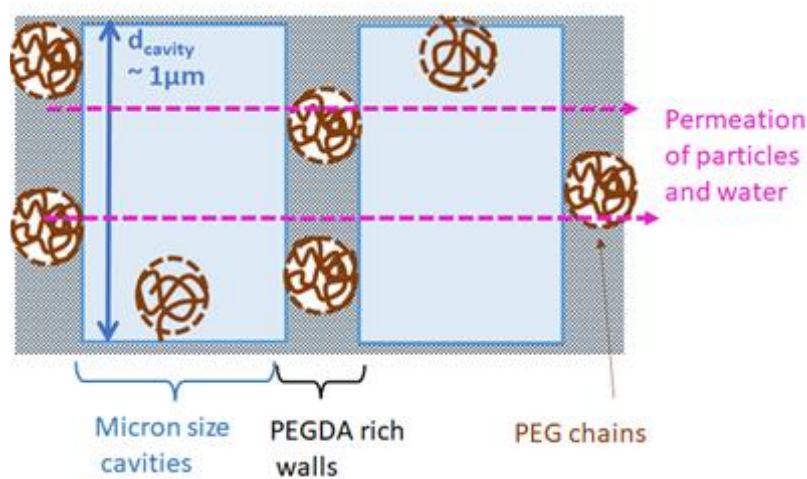

*Figure 9. Schematic propositions of the hydrogel structure that may account for our results. Illustration of the permeation of water and nanoparticles through the PEGDA/PEG hydrogel membranes. The PEG chains trapped in the walls separating the micron sized cavities control the permeation of the particles and water through the hydrogels.*

**CONCLUSION**

We address the filtration of latex particles with different sizes through a composite hydrogel system prepared by photopolymerization of a mixture of PEGDA and large PEG chains of 300 000 g.mol$^{-1}$. By investigating the filtration of nano and microparticles and combining these results with AFM and cryoSEM images we obtain insight into the structure of the PEG/PEGDA hydrogels. We show that particles of 20 nm can permeate through the PEGDA/PEG hydrogels membranes depending on the PEG content and the applied pressure. However, despite the presence of micron size cavities for the PEG containing samples, particles of 100 nm and 1 µm do not permeate through the PEGDA/PEG hydrogel membranes. Moreover, we know from a previous study that the PEG chains are trapped inside the hydrogels and do not get flushed out in the supernatant over time nor during the filtration. Consequently, we suggest that the presence of PEG chains induces local nanoscale defects in the cross-linking of PEGDA-rich walls separating the micron size cavities, that control the permeation of particles and water.

We also investigated the decrease of the water flux in the presence of the latex particles of different sizes. The reduction of the permeation flux in the presence of the particles is probably due to the accumulation of the particles on the surface of the membranes obstructing some of the pores at the surface.

**Supporting Information**

S1. Characterization of latex particles.

S2. Composition of the PEGDA/PEG samples used in this study.

S3. Calibration curves and raw UV/vis data.

S4. Time evolution of permeate volumes of the nano and microparticles through the PEG/PEGDA hydrogels.

S5. Cyclic filtration of NP-100 nm and MP-1 µm

S6. CryoSEM image.

**Acknowledgments**

We gratefully acknowledge Institut Carnot for microfluidics for the financial support during this research project, as well as the Agence Nationale de la Recherche (grants ANR-21-ERCC-0010-01 *EMetBrown,* ANR-21-CE06-0029 *Softer,* ANR-21-CE06-0039 *Fricolas*), and the European Union through the European Research Council (grant ERCCoG-101039103 *EmetBrown*). We also thank the Soft Matter Collaborative Research Unit, Frontier Research Center for Advanced Material and Life Science, Faculty of Advanced Life Science at Hokkaido University, Sapporo, Japan.

# Sieving and clogging in PEG-PEGDA hydrogel membranes

Malak Alaa Eddine, Alain Carvalho, Marc Schmutz, Thomas Salez, Sixtine de Chateauneuf-Randon, Bruno Bresson, Sabrina Belbekhouche, Cécile Monteux

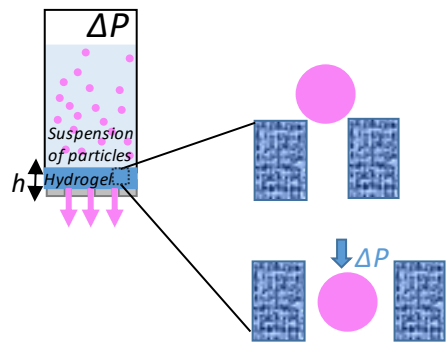